\documentclass[a4paper,UKenglish,cleveref, autoref, thm-restate,final]{lipics-v2021}
\usepackage{xspace}
\usepackage{mathtools}
\usepackage{mathrsfs}
\usepackage{stmaryrd}
\usepackage[obeyFinal]{todonotes}
\usepackage{subcaption}

\hideLIPIcs
\nolinenumbers

\usetikzlibrary{automata, positioning, arrows}
\tikzset{->,>=latex,node distance=1.5cm and 1.5cm,initial text=$ $,accepting/.style = accepting by arrow,auto}

\def \margcol[#1]#2{}

\newcommand{\nl}[1]{\margcol[red!50!black]{Nathan : #1}}

\newcommand{\Ac}{\mathcal{A}}
\newcommand{\Bc}{\mathcal{B}}

\newcommand{\Dc}{\mathcal{D}}

\newcommand{\evalm}[1]{\sem{#1}(m)}

\newcommand{\sem}[1]{\llbracket #1\rrbracket}

\renewcommand{\epsilon}{\varepsilon}

\newcommand{\K}{\mathbb{K}}
\newcommand{\N}{\mathbb{N}}
\newcommand{\Z}{\mathbb{Z}}
\newcommand{\Q}{\mathbb{Q}}

\newcommand{\ninf}{{\N_\infty}}

\newcommand{\set}[1]{\left\{#1\right\}}

\newcommand{\eps}{\epsilon_M}

\newcommand{\invlab}[1]{{\lambda^{-1}_{#1}}}

\newcommand{\waut}[1]{{\pi_{#1}}}
\newcommand{\wautac}{\waut{}}

\newcommand{\efa}[1]{#1_{\lambda_\Ac,\pi_\Ac}}
\newcommand{\psplit}[1]{{P_{#1}}}

\newcommand{\lang}[1]{\mathcal{L}(#1)}
\newcommand\Lang{\mathcal{L}} 
\newcommand{\ax}{\textbf{Ax.}}
\newcommand{\axref}[1]{\ax\ref{#1}}
\newcommand{\col}[1]{I(#1)}
\newcommand{\una}[1]{\phi(#1)}
\newcommand{\indic}[1]{\chi_{#1}}

\newcommand{\ovl}[1]{\overline{#1}}

\newcommand\series{\K\langle\!\langle M \rangle\!\rangle}

\newcommand\amb[2]{\mathrm{amb}(#1,#2)}

\newcommand{\fim}{\mathcal{I}}
\newcommand{\mydag}[1]{{#1}^\dagger}

\newcommand{\bmark}{{\vdash}}
\newcommand{\emark}{{\dashv}}

\title{Weighted Automata and Expressions over Pre-Rational Monoids}
\newcommand{\lis}{Aix Marseille Univ, CNRS, LIS, Marseille, France}

\newcommand{\myauthor}[3]{\author{#1 #2}{\lis}{#1.#2@univ-amu.fr}{https://orcid.org/#3}}
\myauthor{Nicolas}{Baudru}{0000-0002-1333-3432}{}
\myauthor{Louis-Marie}{Dando}{0000-0002-0199-8883}{}
\myauthor{Nathan}{Lhote}{0000-0003-3303-5368}{}
\myauthor{Benjamin}{Monmege}{0000-0002-4717-9955}{}
\author{Pierre-Alain Reynier}{\lis}{pierre-alain.reynier@univ-amu.fr}{}{}
\author{Jean-Marc Talbot}{\lis}{jean-marc.talbot@univ-amu.fr}{}{}

\authorrunning{N. Baudru, L.-M. Dando, N. Lhote, B. Monmege, P.-A. Reynier, J.-M. Talbot} 

\Copyright{Nicolas Baudru, Louis-Marie Dando, Nathan Lhote, Benjamin Monmege, Pierre-Alain Reynier, and Jean-Marc Talbot
}

\keywords{Weighted Automata and Expressions,
  Inverse Monoids, Two-Way Automata} 

\EventEditors{Florin Manea and Alex Simpson}
\EventNoEds{2}
\EventLongTitle{30th EACSL Annual Conference on Computer Science Logic (CSL 2022)}
\EventShortTitle{CSL 2022}
\EventAcronym{CSL}
\EventYear{2022}
\EventDate{February 14--19, 2022}
\EventLocation{G\"{o}ttingen, Germany (Virtual Conference)}
\EventLogo{}
\SeriesVolume{216}
\ArticleNo{22}

\funding{Supported by the DeLTA ANR project
  (ANR-16-CE40-0007)}

\ccsdesc{Theory of computation~Formal languages and automata theory}

\begin{document}
\maketitle
\begin{abstract}
The Kleene theorem establishes a fundamental link
between automata and expressions over the free monoid.
Numerous generalisations of this result exist in the literature.
Lifting this result to a weighted setting
has been widely studied. Moreover, different monoids can be considered:
for~instance,  two-way automata, and even tree-walking automata,
can be described by expressions using the free inverse monoid.
In the present work, we aim at combining both research directions
and consider weighted extensions of automata and expressions
over a class of monoids that we call pre-rational,
generalising both the free inverse monoid and graded monoids.
The presence of idempotent elements in these pre-rational monoids
leads in the weighted setting to consider infinite sums.
To handle such sums, we will have to restrict ourselves
to rationally additive semirings.
Our main result is thus a generalisation of the Kleene theorem
for pre-rational monoids and rationally additive semirings.
As a corollary, we obtain a class of expressions equivalent
to weighted two-way automata, as well as one for tree-walking automata.

\end{abstract}

\section{Introduction}

Automata are a convenient tool for algorithmically processing regular
languages. However, when a short and human-readable description is
required, regular expressions offer a much more proper formalism. When
it comes to weighted automata (and transducers as a special case), the
Kleene-Schützenberger theorem \cite{Sch61} relates weighted languages
defined by means of such automata on one side, and rational series on
the other side. Unfortunately, such expressions seem to fit mainly for
one-way machines. Indeed, when it comes to two-way machines, finding
adequate formalisms for expressions is not easy (see, e.g.,
\cite{Lom16} where Lombardy introduces a new matricial product
to faithfully represent two-way automata).\todo{BM: comme le suggérait
  LM, j'ai ajouté la
  parenthèse pour répondre au reviewer 3 (can
  you give an intuition or at least a reference to justify "... is not
  easy.")}

Two-way automata have been studied in the setting of the Boolean
semiring in~\cite{DJ15}. In this work, Janin and Dicky consider a
fragment of the 
free inverse monoid called overlapping tiles.
They show that runs of a two-way automaton can be described as
a recognizable language of overlapping tiles,
which are words enriched with a starting and an ending position.
Hence, thanks to the Kleene theorem, such two-way runs can be
described as regular expressions (over tiles).

A particular class of weighted automata is that of transducers, where
weights are words on an output alphabet\todo{BM: j'ai ajouté la fin de
  la phrase, comme suggéré par NL, pour répondre au Reviewer 3 (first
  paragraph in page 2 - not clear)}.  For this setting, Alur \emph{et
  al} proposed in~\cite{DBLP:conf/csl/AlurFR14} a formalism to
describe word transformations given as a deterministic streaming
string transducer, a model equivalent with deterministic (or
unambiguous) two-way transducers~\cite{FR-SIGLOG16}.  This formalism
is based on some operators defining basic transformations that are
composed to define the target transformation. An alternative
construction of these expressions starting directly from two-way
unambiguous transducers has been proposed
in~\cite{DBLP:journals/ijfcs/BaudruR20}. These expressions have also
been extended to run on infinite words
\cite{DBLP:conf/lics/DaveGK18}. The general case of non-deterministic
two-way transducers is much more challenging~\cite{phdGuillon}, as
these machines may admit infinitely many accepting runs on an input
word.  While this general case is still open (meaning that no
equivalent models of expressions are known), a solution has been
proposed for the particular case where both input and output alphabets
are unary~\cite{ChoffrutG14}.

For a further weighted generalisation, the ability to sum values computed by
different runs on the same input structure (no matter if it is a word,
a tree or even a graph) is also crucial in terms of
expressiveness. However, not all weighted two-way automata (or
weighted one-way automata with $\epsilon$-transitions) are valid:
indeed, as these machines may have infinitely many runs over a single
input, it may be the case that the automaton does not provide any
semantics for such inputs, infinite sums being not guaranteed to
converge. To overcome this issue, additional properties are required
over the considered semiring: for instance, \emph{rationally additive
  semirings} \cite{EsKu02} allow one to define valid non-deterministic
two-way automata \cite{LoSa13}.

Our initial motivation was to elaborate on the approach proposed by
Janin and Dicky in the setting of weighted languages. As already said,
the main ingredient of their approach is to consider the free inverse
monoid as an input structure. However, going one step further, we
consider a generalisation, namely \emph{pre-rational monoids}. These
are monoids $M$ such that for all finite alphabets $A$ and for all
morphisms from the free monoid $A^*$ to $M$, the pre-image of $m\in M$
is a rational language of $A^*$. This class of monoids contains, in
particular, the free inverse monoid. After introducing the monoids and
semirings of interest in Section~\ref{sec:definitions}, we present
our main contributions, which hold for pre-rational
monoids and rationally additive~semirings:
\begin{enumerate}
\item We prove in Section~\ref{sec:WA} that
  all weighted automata are valid.
\item We introduce in Section~\ref{sec:WE} a syntax for weighted
  expressions and show that 
   the semantics of these expressions is always
  well-defined. 
\item We prove in Section~\ref{sec.kth} a Kleene-like theorem stating
  that
   weighted automata and weighted expressions
  define the same~series.
\item We deal with the particular case of unambiguous automata and
  expressions in Section~\ref{sec:unambiguous}. More precisely, our
  conversions are shown to preserve the ambiguity, meaning that an
  element of the monoid ``accepted'' $k$ times by a weighted automaton
  can be ``decomposed'' in $k$ different ways by the weighted expression we
  obtain, and vice~versa\todo{new sentence}.
\item In Section~\ref{sec:fim}, we apply our results on two-way word
  automata and tree-walking automata which can be viewed as part of
  the free inverse monoids (which are pre-rational) and show how
  expressions are quite natural to write via a variety of examples. As
  a corollary, we obtain a formalism of expressions equivalent to
  non-deterministic two-way transducers. To do so, we use the
  unambiguity result presented in the previous section\todo{new
    sentence}.
\end{enumerate}

Our results can be understood as a trade-off between the
generality of the monoid and that of the semiring. 
Indeed, instead of rationally additive semirings, one could
have considered \emph{continuous semirings} in which all infinite sums
are well-defined. On such semirings, weighted automata are valid on
all input monoids \cite{Saka09}. However, our framework
allows one to consider semirings that are not continuous, and as a consequence we
have to restrict in this case the input monoid. On the other end of
the spectrum, restricting oneself to graded monoids (as also done
in~\cite{Saka09}) allows one to consider any semiring, since only
finite sums are then involved. However, the free inverse monoid is a
typical example of non-graded monoid.

\section{Monoids and semirings}
\label{sec:definitions}

We recall that a \emph{monoid} $(M,\cdot,\eps)$ is given by a set $M$
and an associative product $\cdot$ with~$\eps$ as neutral element.
For our purpose, we consider special classes of monoids:


%


\begin{definition}
  A monoid $(M,\cdot,\eps)$ is \emph{pre-rational} if for every finite
  alphabet~$A$, for every morphism $\mu\colon A^*\rightarrow M$, and
  for every $m\in M$, the language $\mu^{-1}(m)\subseteq A^*$ is
  rational.
\end{definition}

Many natural examples of monoids are pre-rational: the free monoid
$(A^*, \cdot, \varepsilon)$ over a finite alphabet $A$, the natural
monoid $(\N, +, 0)$, and even the one completed with an infinite
element $(\N\cup \{+\infty\}, +, 0)$. Other examples, of particular
interest in this article, are \emph{free inverse monoids} that we
study in Section~\ref{sec:fim}.
Another non-trivial example of pre-rational monoid is
$(\{L \subseteq A^* \mid \varepsilon\in L\}, \cdot, \{\varepsilon\})$,
with $A$ a finite alphabet (see Appendix~\ref{app:prerat} for a proof).
In contrast, a typical example of monoid that is not pre-rational is
the free group generated by one element, or $(\Z,+,0)$
equivalently. For instance, given the morphism
$\mu\colon \set{a,\bar a}^*\rightarrow \Z$ mapping $a$ to $1$ and
$\bar a$ to ${-}1$, then
$\mu^{-1}(0)=\set{w\in \set{a,\bar a}^*|\quad |w|_a=|w|_{\bar a}}$
which is not rational.

Showing pre-rationality might sometimes be challenging, since
considering arbitrary alphabets and arbitrary morphisms is not really
convenient. An easier definition is however possible for monoids $M$
that are generated by a finite family $G = \{g_1,\ldots,g_n\}$ of
generators. In this case, consider the canonical morphism $\varphi$
from the free monoid $G^*$ (considering generators as letters) to $M$,
that consists in evaluating the sequence of generators in $M$. Then,
$M$ is pre-rational if and only if for all $m\in M$, the language
$\varphi^{-1}(m)\subseteq G^*$ is rational. Pre-rationality is then
easier to check, and this, without much of a restriction:
the automata and expressions we will consider thereafter only use a finite
set of elements of the monoid as atoms, and we can thus restrict
ourselves to the finitely generated submonoid. An even simpler
sufficient condition for pre-rationality is:

\begin{lemma}\label{lem.pref}
  If every element $m$ of a monoid $M$ has a finite number of
  \emph{prefixes}, i.e.~elements~$p\in M$ such that there exists
  $p'\in M$ with $m=p \cdot p'$, then $M$ is pre-rational.
\end{lemma}
\begin{proof}
  For a finite alphabet $A$ and a morphism $\mu\colon A^*\to M$, and
  an element $m\in M$, with $\{m_1,\ldots,m_n\}$ as finite set of
  prefixes, we can build a finite automaton reading letters of~$A$
  and, after having read a word $w\in A^*$, storing the current element
  $\mu(w)$, if it is a prefix of~$m$ (going to a non-accepting sink
  state otherwise). This automaton can then be used to
  recognise $\mu^{-1}(m)$, by starting in the prefix $\eps$ and
  accepting in the prefix $m$.
\end{proof}

This allows us to easily show that all finitely generated graded
monoids~\cite{Saka09} (i.e.~monoids~$M$ equipped with a gradation
$\varphi\colon M\to \N$ such that $\varphi(m)=0$ only if $m=\eps$,
and $\varphi(mn) = \varphi(m)+\varphi(n)$ for all $m,n\in M$)
are pre-rational. Indeed,
the gradation ensures that each element $m\in M$ can have only
a finite number of prefixes \cite[Chap.~III, Cor.~1.2,p.384]{Saka09},
allowing us to apply the previous lemma.
However, notice that the condition in Lemma~\ref{lem.pref} is not a
necessary one: $(\N\cup\{+\infty\}, +, 0)$ does not fulfil the
condition, since $+\infty$ has infinitely many factors, while it is
indeed pre-rational.  


A \emph{semiring} $(\K,+,\times,0,1)$ is an algebraic structure such
that $(\K,\times, 1)$ is a monoid, $(\K,+, 0)$ is a commutative
monoid
, the product $\times$ distributes over the sum $+$ 
, and $0$ is absorbing
for~$\times$
.
Once again, we consider special classes of semirings, 
introduced in \cite{EsKu02}:
\begin{definition}
  A semiring $(\K,+,\times,0,1)$ is \emph{rationally additive} if it
  is equipped with a partial operator defining sums of countable
  families, associating with some infinite families
  $(\alpha_i)_{i\in I}$, with $I$ at most countable, an element
  $\sum_{i\in I} \alpha_i$ of $\K$ such that for all families
  $(\alpha_i)_{i\in I}$:
  \begin{enumerate}[\ax 1]
  \item\label{ax1} If $I$ is finite, the value
    $\sum_{i\in I} \alpha_i$ exists and coincides with the usual sum
    in the semiring.
  \item \label{ax2} For each $\alpha\in\K$,
    $\sum_{n=0}^{\infty}\alpha^n$ exists.
  \item \label{ax3} If $\sum_{i\in I}\alpha_i$ exists and
    $\beta\in\K$, then $\sum_{i\in I}\beta\alpha_i$ and
    $\sum_{i\in I}\alpha_i\beta$ exist, and are respectively equal to
    $\beta(\sum_{i\in I}\alpha_i)$ and $(\sum_{i\in I}\alpha_i)\beta$.
  \item\label{ax4} Let $I$ be the disjoint union of $(I_j)_{j\in J}$
    with $J$ at most countable.  If for all $j\in J$,
    $r_j=\sum_{i\in I_j}\alpha_i$ exists, and if $r=\sum_{j\in J}r_j$
    exists, then $\sum_{i\in I}\alpha_i$ exists and is equal to $r$.
  \item\label{ax5} Let $I$ be the disjoint union of $(I_j)_{j\in J}$
    with $J$ at most countable.  If $s=\sum_{i\in I}\alpha_i$ exists,
    and for all $j\in J$, $r_j=\sum_{i\in I_j}\alpha_i$ exists, then
    $\sum_{j\in J}r_j$ exists and is equal to $s$.
  \end{enumerate}
\end{definition}

Examples of rationally additive semirings are the Boolean semiring,
natural semirings over positive rationals or reals
$(\Q_+\cup\{\infty\}, +,\times, 0, 1)$\footnote{All infinite
  sums of elements in $\Q_+$ do not converge towards a rational number
  or $+\infty$, but all \emph{geometric sums} do. In particular, this
  semiring is not \emph{continuous} (see~\cite[Chap.~III,
  Sec.~5]{Saka09}).}, the tropical (or arctic) semiring
$(\Q\cup\set{{-}\infty,{+}\infty},\sup,+,{-}\infty, 0)$, the language
semiring over a finite alphabet
$(2^{A^*}, \cup, \cdot, \emptyset,\{\varepsilon\})$, the sub-semiring
of rational languages, or distributive lattices.
Throughout this article, $\K$ will denote a rationally additive
semiring.

Let us state a few useful properties of rationally additive semirings.
The \emph{support} of a family $(\alpha_i)_{i\in I}$ is the set
$\{i\in I\mid \alpha_i\neq 0\}$ of indices of non-zero elements.

\begin{lemma}[\cite{EsKu02}]\label{lem.zsum}
  Let $(\alpha_i)_{i\in I}$ be a countable family in $\K$, of support
  $J$. Then, $\sum_{i\in I}\alpha_i$ exists if and only if
  $\sum_{i\in J}\alpha_i$ exists, and when these sums exist, they are
  equal.
\end{lemma}

\begin{lemma}\label{lem.linsum}
  Let $(\alpha_i)_{i\in I}$ and $(\beta_i)_{i\in I}$ be two countable families
  of $\K$ of disjoint supports, i.e.~for all $i\in I$, $\alpha_i=0$ or
  $\beta_i=0$ (or both).  If $\sum_{i\in I}\alpha_i$ and $\sum_{i\in I}\beta_i$
  exist, then $\sum_{i\in I}(\alpha_i+\beta_i)$ exists and is equal to
  $(\sum_{i\in I}\alpha_i)+(\sum_{i\in I}\beta_i)$.
\end{lemma}
\begin{proof}
  Let $J_\alpha$ and $J_\beta$ be the support of the families
  $(\alpha_i)_{i\in I}$ and $(\beta_i)_{i\in I}$, and
  $J_0=J\setminus(J_\alpha\cup J_\beta)$.  If $\sum_{i\in I}\alpha_i$
  and $\sum_{i\in I}\beta_i$ exist,
  $\sum_{i\in I}\alpha_i+\sum_{i\in I}\beta_i$ exists, and by
  Lemma~\ref{lem.zsum}, is equal to
  $\sum_{i\in J_\alpha}\alpha_i+\sum_{i\in J_b}\beta_i$.  Since the
  supports are disjoint, this is equal to
  $\sum_{i\in J_\alpha}(\alpha_i+\beta_i)+\sum_{i\in
    J_b}(\alpha_i+\beta_i)$.  By definition of $J_0$,
  $\sum_{i\in J_0}(\alpha_i+\beta_i)$ exists and is equal to 0.
  Therefore, $\sum_{i\in I}\alpha_i+\sum_{i\in I}\beta_i$ is equal to
  $\sum_{i\in J_\alpha}(\alpha_i+\beta_i)+\sum_{i\in
    J_b}(\alpha_i+\beta_i)+\sum_{i\in J_0}(\alpha_i+\beta_i)$.
  \axref{ax4} allows us
  to conclude.
\end{proof}

\begin{lemma}\label{lem.permsum}
  Let $(\alpha_{i,j})_{(i,j)\in I\times J}$ be a countable family of
  elements of $\K$, such that $\alpha_{i,J}=\sum_{j\in J}\alpha_{i,j}$ exists
  for all $i\in I$, and $\alpha_{I,j}=\sum_{i\in I}\alpha_{i,j}$ exists for all
  $j\in J$.  Then, $\sum_{i\in I}\alpha_{i,J}$ exists if and only if
  $\sum_{j\in J}\alpha_{I,j}$ exists, and when these sums exists, they are
  equal.
\end{lemma}
\begin{proof}
  Immediate by~\axref{ax4} and~\axref{ax5}.
\end{proof}

\section{Series and Weighted Automata}\label{sec:WA}

A \emph{$\K$-series over $M$} is a mapping $s\colon M\rightarrow \K$
associating a weight $s(m)$ with each element $m$ of the monoid.  The
set of all such series is denoted by $\series$.  Notice that the
pointwise sum of two series $s_1$ and $s_2$, defined for all $m\in M$
by $(s_1+s_2)(m) = s_1(m)+s_2(m)$, is a series. However, the Cauchy
product $s_1\cdot s_2$ mapping $m$ to the possibly infinite sum
$\sum_{m_1m_2=m} s_1(m_1)\times s_2(m_2)$ might not
exist\footnote{Here and in the following, $\sum_{m_1m_2=m}$ is
  the sum over all pairs $(m_1,m_2)\in M^2$ such that $m_1m_2=m$.}.
We define two canonical injections: $M\rightarrow \series$ which maps
$m$ to the characteristic function of $m$ (mapping $m$ to $1$ and the other elements from $M$ to $0$), 
and $\K\rightarrow \series$ which maps $k$ to the function mapping the neutral element $\eps$ of
$M$ to $k$ and all other values to $0$. For this reason, we often
abuse notations and consider $\K$ and $M$ as subsets of
$\series$. 

We now introduce the notion of weighted automata 
we consider in this article:
weights are taken from a rationally additive semiring
$\K$ and \emph{labels} from a pre-rational monoid $M$.
\begin{definition}
  A \emph{$\K$-automaton over $M$}, or simply a \emph{weighted
    automaton}, is a tuple~$\Ac=(Q,I,\Delta,F)$, with $Q$ a finite set
  of states, $I\subseteq Q$ the set of initial states,
  $\Delta\subseteq Q\times M\times \K\times Q$ the finite set of
  transitions each equipped with a label in $M$ and a weight in $\K$,
  and $F\subseteq Q$ the set of final states.  
\end{definition}

We introduce two mappings $\lambda_\Ac$ and $\pi_\Ac$ that extract the
label and the weight of a transition, that we can extend to morphisms
from $\Delta^*$ to $M$ and the multiplicative monoid of $\K$,
respectively.  A \emph{run} of $\Ac$ is then a sequence $w$ of
transitions $(p_i,m_i,k_i,q_i)_{1\leq i \leq n}$ such that for all
$i$, $q_i=p_{i+1}$.  The \emph{label} of a run is given by
$\lambda_\Ac(w)$; its weight is $\pi_\Ac(w)$.
The run is said to be
\emph{accepting} if $p_1\in I$ and $q_{n}\in F$. We let
$R_\Ac\subseteq \Delta^*$ 
denote the rational language of all accepting runs.
The \emph{semantics} of $\Ac$ is the series $\sem{\Ac}$
such that for all $m\in M$, the weight $\sem\Ac(m)$ is
the sum of the weights of accepting
runs that are labelled by $m$, if the (potentially infinite) sum
exists: $\sem\Ac(m)=\sum_{w\in R_\Ac\cap
  \lambda_\Ac^{-1}(m)}\pi_\Ac(w)$. 

The automaton $\Ac$ is called \emph{valid} if the sum in $\sem\Ac(m)$
exists for all $m\in M$. Instead of enforcing properties on the
automata for them to be valid, we ensure their validity by combining the
rational additivity of $\K$ and the pre-rationality of $M$. The
crucial technical property considers the special case of the monoid of
strings $A^*$ over a finite alphabet $A$. We then lift the result using
pre-rationality.
For a language $L\subseteq A^*$ and a semiring $\K$, we denote by
$\indic{L}\in\K\langle\!\langle A^*\rangle\!\rangle$ its
\emph{characteristic series} in $\K$, defined for all $w\in A^*$ as
$\indic{L}(w)=1$ if $w\in L$, and 0
otherwise. 
By Lemma~\ref{lem.zsum}, we have that for all series $s$ over $A^*$,
\begin{equation}
  \text{$\sum_{w\in L}s(w)$ is defined iff $\sum_{w\in A^*}s(w)\indic{L}(w)$
    is defined, and then these sums are equal.}\label{eq:caracteristic}
\end{equation}

\begin{lemma}\label{lem.exists}
  For every finite alphabet $A$, morphism $\pi\colon A^*\to \K$, and
  rational language $L\subseteq A^*$, the sum $\sum_{w\in L}\pi(w)$
  exists.
\end{lemma}
\begin{proof}
  The proof is by induction on rational languages, denoted by
  unambiguous regular expressions~\cite{BooEve71}.
  Indeed, all rational languages can be obtained
  by closing the set of finite languages by
  the operations of disjoint unions, unambiguous concatenations
  (the concatenation $L_1\cdot L_2$ is unambiguous
  when each word $w$ of $L_1\cdot L_2$ can be uniquely decomposed
  as $w=w_1\cdot w_2$ with $w_1\in L_1$ and $w_2\in L_2$),
  and unambiguous Kleene stars (the Kleene star $L^*$ is
  unambiguous when each word $w\in L^*$ can be
  uniquely decomposed as $w=w_1 \cdots w_n$ with $n\in \N$ and
  $w_i\in L$ for all $i$).
  Please note that for convenience,
  the sentences ``$A=B$'' should be
  read as ``$B$ exists and is equal to
  $A$''.  
  
  First, for finite languages $L$, the sum $\sum_{w\in L}\pi(w)$
  exists, by \axref{ax1}. 
  In the case where $L$ is the disjoint union of two languages $L_1$
  and $L_2$, such that $\sum_{w\in L_1}\pi(w)$ and
  $\sum_{w\in L_2}\pi(w)$~exist, 
    \begin{align*}
      \sum_{w\in L_1}\pi(w)+
      \sum_{w\in L_2}\pi(w)        
      &=\sum_{w\in A^*}\pi(w)\indic{L_1}(w)+
      \sum_{w\in A^*}\pi(w)\indic{L_2}(w)
      &\text{(by \ref{eq:caracteristic})}\\
      &=\sum_{w\in A^*}(\pi(w)\indic{L_1}(w)+\pi(w)\indic{L_2}(w))
      &\text{(by Lemma~\ref{lem.linsum})}\\
      &=\sum_{w\in A^*}\pi(w)\indic{L_1\cup L_2}(w)
      &\text{(disjoint union)}\\
      &=\sum_{w\in L_1\cup L_2=L}\pi(w). 
    \end{align*}

    If $L$ is the unambiguous concatenation of two languages $L_1$ and
    $L_2$ such that $\sum_{u\in L_1}\pi(u)$ and
     $\sum_{v\in L_2}\pi(v)$ exist,
    then 
    \begin{align*}
      \Big(\sum_{u\in L_1}\pi(u)
      \Big)\times \Big(
      \sum_{v\in L_2}\pi(v)\Big)
      &=\sum_{u\in L_1}\Big(\pi(u)
      \times
      \sum_{v\in L_2}\pi(v)\Big)
      &\text{(by \axref{ax3})}\\
      &=\sum_{u\in L_1}
      \sum_{v\in L_2}\pi(u)\pi(v)
      &\text{(by \axref{ax3})}\\
      &=\sum_{(u,v)\in L_1\times L_2}\pi(u)\pi(v)
      &\text{(by \axref{ax4})}\\
      &=\sum_{(u,v)\in L_1\times L_2}\pi(uv)
      &\text{($\pi$ is a morphism).}
    \end{align*}
    Moreover, by unambiguity,
    there exists a bijection from the pairs of $L_1\times L_2$
    to the words of the concatenation $L_1\cdot L_2$
    sending $(u,v)$ to $uv$. Bijections on the support
    of families conserve the summability property
    by~\cite[Proposition 3]{EsKu02},
    therefore $\sum_{w\in L}\pi(w)$ exists
    (and is equal to $\sum_{(u,v)\in L_1\times L_2}\pi(uv)$).
    
    Finally, suppose that $L$ is the unambiguous Kleene star $L_1^*$, and
    $\sum_{w\in L_1}\pi(w)$ exists. In particular, for all $n\in \N$,
    the iterated concatenation $L_1^n$ is unambiguous, and thus, with
    a straightforward induction using the previous case,
    $\smash[b]{\sum_{w\in L_1^n}}\pi(w)$
    exist and we have
    \[\Big(\sum_{w\in L_1}\pi(w)\Big)^n=\sum_{w\in L_1^n}\pi(w).\]
    By \axref{ax2}, $\sum_{n=0}^\infty\Big(\sum_{w\in L_1}\pi(w)\Big)^n$
    exists, and by~\eqref{eq:caracteristic}, we have:
    \begin{align*}
      \sum_{n=0}^\infty
      \Big(\sum_{w\in L_1}\pi(w)\Big)^n
      &=\sum_{n=0}^\infty
        \sum_{w\in L_1^n}\pi(w)=\sum_{n=0}^\infty
        \sum_{w\in A^*}\pi(w)\indic{L_1^n}(w).
    \end{align*}
    By unambiguity, for all
    $w\in A^*$, the infinite sum
    $\sum_{n=0}^\infty \pi(w)\indic{L_1^n}(w)$ has finite support (at
    most 1 non-zero element) and therefore exists (by
    Lemma~\ref{lem.zsum}). By Lemma~\ref{lem.permsum}, we deduce that
    \begin{align*}
      \sum_{n=0}^\infty
        \sum_{w\in A^*}\pi(w)\indic{L_1^n}(w)
      &=\sum_{w\in A^*}
        \sum_{n=0}^\infty
        \pi(w)\indic{L_1^n}(w)=\sum_{w\in A^*}\pi(w)
        \sum_{n=0}^\infty
        \indic{L_1^n}(w)
      &\text{(by \axref{ax3})}\\
      &=\sum_{w\in A^*}\pi(w)
        \indic{L_1^*}(w)
      &\makebox[2.2cm][r]{\text{(by unambiguity)}}\\
      &=\sum\limits_{w\in L}\pi(w). \tag*{\qedhere}
    \end{align*}
\end{proof}

From this result, to have a sufficient
condition for validity
we only need to have sums over rational languages,
hence our requirement that $M$ is pre-rational.

\begin{theorem}\label{theo.automata}
  If $M$ is a pre-rational monoid, then every $\K$-automaton $\Ac$
  over $M$ is valid, i.e. $\sem{\Ac}(m)$ exists for all $m\in M$.
\end{theorem}
\begin{proof}
  Since $M$ is pre-rational, the morphism $\lambda_\Ac$ is such that
  for all $m\in M$, $\lambda_\Ac^{-1}(m)$ is a rational
  language. Therefore, so is the language
  $R_\Ac\cap \lambda_\Ac^{-1}(m)$ of accepting runs that are labelled
  by the element $m$. Lemma~\ref{lem.exists} gives that
  $\sem{\Ac}(m)=\sum_{w\in R_\Ac\cap \lambda_\Ac^{-1}(m)} \pi_\Ac(w)$
  exists.
\end{proof}

Together with reasonable assumptions on computability
for $\K$ and $M$, this also gives a procedure to
evaluate the weight $\sem\Ac (m)$. Notice that this is a priori non-trivial, since
it involves an infinite sum. 
We say that $M$ is \emph{effectively} pre-rational
if for all morphisms~$\mu\colon A^* \rightarrow M$ and elements~$m\in M$, 
one can compute a representation of the rational language~$\mu^{-1}(m)$.
We say that $\K$ is \emph{computable} if internal operations (finite
sums and products) of $\K$ are computable,
as well as Kleene star (geometric sum).
Observe that we do not require computability of arbitrary
infinite sums, but only geometric ones. 
\begin{proposition}
  If $M$ is effectively pre-rational and $\K$ is computable, then for
  all $\K$-automata $\Ac$ over $M$ and all elements $m\in M$, one can
  compute $\sem\Ac (m)$.\todo{Reviewer 3: what is the complexity
    required?  LM: ça dépend de "effectively pre-rat", du coup je ne
    vois pas comment reformuler.  }\nl{ça dépend aussi de la complexité des opérations dans $\K$. On peut dire que la complexité de la première étape (qui utilise la pré-rationalité effective) et de la dernière étape (évaluer une $\K$-expression) dépendent respectivement de $M$ et de $\K$, mais que l'étape du milieu est purement syntactique la complexité est la même que dans le cas des expressions de Kleene}
\end{proposition}


\begin{proof}
  By assumption of pre-rationality, the language $\lambda_\Ac^{-1}(m)$
  is rational. 
  Moreover, by effectiveness, we can let $\Dc_m$ be a deterministic
  automaton that recognises $\lambda_\Ac^{-1}(m)$. The~$\K$-automaton
  $\Ac_m$ obtained by considering the product of $\Ac$ and $\Dc_m$
  (with respect to the alphabet $\Delta$ of transitions of $\Ac$)
  thus restricts the runs of $\Ac$ to the ones labelled by~$m$. By
  removing all labels (replacing them by $\eps$), we obtain a
  $\K$-automaton that associates with the element $\eps$ the weight
  $\sem{\Ac_m}(\eps)=\sem{\Ac}(m)$. Applying classical translations
  from automata to regular expressions such as state-elimination
  algorithms yields an expression equivalent to $\sem\Ac (m)$. This
  expression involves sum and product in $\K$, as well as Kleene
  star, which can be computed in $\K$.
\end{proof}




\section{Weighted Expressions}\label{sec:WE}\label{sec:wexp}

We now introduce the formalism of weighted expressions to generate
$\K$-series over a monoid~$M$.

\begin{definition}
  The set of \emph{$\K$-expressions over $M$}, or simply
  \emph{weighted expressions}, is generated by the grammar (with
  $k\in\K$ and $m\in M$):
  \[W::= k\mid m\mid W+W\mid W\cdot W\mid W^*.\]
\end{definition}
Expressions $k$ and $m$ are said to be \emph{atomic}. We call
\emph{subexpressions} of $W$ all the weighted expressions appearing
inside $W$: for instance, the subexpressions of $W=(2\cdot a + b)^*$
are $2$, $a$, $b$, $2\cdot a$, $2\cdot a+b$, and $W$. \todo{BM: deux
  phrases et une note de bas de page nouvelles} To define the
semantics of weighted expressions, we will use a sum operator over
infinite families. As the semiring $\K$ is supposed to be rationally
additive, some of these infinite sums exist, some others do
not\footnote{In the rationally additive semiring
  $(\Q_+\cup\{\infty\}, +,\times, 0, 1)$, the infinite sum
  $\sum_{i\in \N} 1/i!$ does not exist, since it converges to the
  non-rational real number $\mathrm e$.}. Then, the semantics of a
weighted expression $W$ is the series $\sem W\in\series$ defined
inductively as follows:
\begin{itemize}
\item $\sem{k}$ is the series mapping $\eps$ to $k$ and other
  elements to $0$;
\item $\sem{m}$ is the characteristic series of $m$;
\item $\sem{U+V}=\sem{U}+\sem{V}$;
\item for all $m\in M$,
  $\sem{U\cdot V}(m)=\sum_{m_1m_2=m}\sem{U}(m_1)\times \sem{V}(m_2)$
  if the sum exists;
\item for all $m\in M$,
  $\sem{W^*}(m)=\sum_{n=0}^{\infty} \sem {W^n}(m)$ if the sum exists
  (with $W^n$ the expression inductively defined by $1$ if $n=0$ and
  $W\cdot W^{n-1}$ otherwise).
\end{itemize}

The last two cases, defining the semantics of the concatenation (or
Cauchy product) of two weighted expressions, and the Kleene star of a
weighted expression, are subject to the existence of the infinite
sums: we say that a weighted expression is \emph{valid} when its
semantics exists for all $m\in M$ (as well as the semantics of all its
subexpressions, in particular).

More usual regular expressions are recovered by considering the
Boolean semiring and the monoid $A^*$ over a finite alphabet $A$: in
the following, such expressions are called \emph{Kleene expressions},
and denoted by letters $E, F, G$, while weighted expressions are
denoted by letters~$U, V, W$.
Notice that Kleene expressions are valid, as expected, since the
infinite \emph{sum} (i.e.~disjunction in the Boolean semiring) is
always defined in this case.  Their semantics $\sem E$ is the
characteristic series of the language $\Lang(E)$ classically
associated with such a regular expression: alternatively, we can see
$\Lang(E)$ as the support of $\sem E$ (all words $w\in A^*$ such that
$\sem E(w)$ is true). For any other semiring
$\K$, we let $\indic{E}$ be the characteristic function of the
language of~$E$ to the semiring $\K$, i.e.~a shortcut notation for the
series $\indic{\Lang(E)}\in\K\langle\!\langle A^*\rangle\!\rangle$
defined in Section~\ref{sec:WA}.

We shall see that thanks to our hypothesis of $\K$ being rationally
additive, and restricting ourselves to pre-rational monoids, all
weighted expressions are valid:

\begin{theorem}\label{thm.expok}
  Let $\K$ be a rationally additive semiring,
  and $M$ be a pre-rational monoid.
  
  Every $\K$-expression $W$ over $M$ is valid,
  \emph{i.e.} the semantics $\sem W(m)$ exists for all $m\in M$.
\end{theorem}

Notice that this theorem relies on both its assumptions on $M$ and
$\K$:
\begin{itemize}
\item If $M$ is not pre-rational, then the expressions may not be
  valid. For instance, consider $M$ to be the free group generated by
  a single element $a$ (with $a^{-1}$ its inverse in the free
  group), and $\K$ be the semiring of rational languages
  over the alphabet $\{A,B\}$.
  Then, the expression $(a \cdot \{A\} + a^{-1} \cdot
  \{B\})^*$ would associate with the element $\eps$ of $M$ the language
  of words over $\{A,B\}$ having as many $A$'s than $B$'s, which is not
  rational, and thus not a member of $\K$.
\item If $\K$ is not rationally additive, then the expressions may not
  be valid. For instance, considering the semiring
  $(\Q, +, \times, 0, 1)$, and the (pre-rational) trivial monoid
  $\{\eps\}$, the expression $W = (-1)^*$ gives as a semantics
  $\sem W(\eps) = \sum_{n\in \N}(-1)^n$ that is the archetypal
  diverging series in $\Q$.
\end{itemize}

The rest of this section is devoted to
the proof of this theorem. This proof is split into two parts
. We first show that the
validity of a weighted expression obtained by the rewriting of
``letters'' in an \emph{unambiguous} Kleene expression is equivalent
to the existence of sums resembling the ones of
Lemma~\ref{lem.exists}. We then explain how to generate such an
unambiguous Kleene expression from a weighted expression $W$, and
apply the previous result to show the validity of $W$.

More formally, a Kleene expression $E$ (over a monoid $A^*$) is
called unambiguous if for all its subexpressions $E'$:
\begin{itemize}
\item if $E'=F+G$, then $\Lang(F)\cap \Lang(G) = \emptyset$;
\item if $E'=F\cdot G$, then for all $w\in A^*$, there exists at
  most one pair $(w_1,w_2)\in \Lang(F)\times \Lang(G)$ such that
  $w_1w_2=w$;
\item if $E' = F^*$, then for all $w\in A^*$, there exists at most
  one natural number $n$, and one sequence
  $(w_1,w_2,\ldots,w_n)\in (\Lang(F))^n$ such that
  $w_1w_2\cdots w_n = w$.
\end{itemize}

As a direct corollary, for every semiring $\K$,
\begin{itemize}
\item if $E+F$ is unambiguous, then $\indic{E+F}=\indic{E}+\indic{F}$;
\item if $E\cdot F$ is unambiguous, then
  $\indic{E\cdot F}(w)=\sum_{uv=w}\indic{E}(u)\indic{F}(v)$;
\item if $E^*$ is unambiguous, then
  $\indic{E^*}=\sum_{n=0}^{\infty}\indic{E^n}$, this infinite sum
  having indeed a finite support and being thus meaningful in any
  semiring (and formally existing in a rationally additive semiring). 
\end{itemize}

Given two morphisms $\lambda\colon A^*\to M$ and $\pi\colon A^*\to\K$,
we let $E_{\lambda,\pi}$ be the weighted expression obtained from a
Kleene expression $E$ by substituting every letter $a$ appearing in
$E$ by the expression $\lambda(a)\cdot \pi(a)$, and by replacing Booleans
\emph{true} and \emph{false} by elements $1\in \K$ and $0\in \K$.

The next lemma aims at linking the validity of $E_{\lambda,\pi}$ with
the existence of specific infinite sums. The same result is also
fundamental in our later proofs of translations between automata
and expressions in the next section.

\begin{lemma}
  \label{lem.central}
  Let~$E$ be an unambiguous Kleene expression over a free
  monoid~$A^*$, $M$ be a monoid (not necessarily pre-rational), $\K$
  be a rationally additive semiring, $\lambda\colon A^*\to M$ and
  $\pi\colon A^*\to\K$ be two morphisms. Then, $E_{\lambda,\pi}$ is
  valid if and only if for all $m\in M$ and all subexpressions $F$ of
  $E$, the sum $\sum_{\lambda(w)= m}\pi(w)\indic{F}(w)$ exists (where
  the sum is over all words $w\in A^*$ such that $\lambda(w)=m$). In
  this case, for all $m\in M$,
  $\sem{{E}_{\lambda,
      \pi}}(m)=\sum_{\lambda(w)=m}\pi(w)\indic{E}(w)$.
\end{lemma}
 \begin{proof}
  
We show the result by induction over $E$.  The existence and
  manipulation of the various infinite sums must be treated carefully:
  for convenience, the sentence "$A=B$ (by $C$)" should be read "$B$
  exists (by $C$) and is equal to $A$ (also by $C$)".
  
  \begin{itemize}
  \item If $E=a$ with $a\in A$, then
    $E_{\lambda,\pi} = \lambda(a)\cdot 1$, so that for all $m\in M$,
    $\sem{E_{\lambda,\pi}}(m)$ is equal to
    $\sum_{m_1m_2 = m} \sem{\lambda(a)}(m_1) \times \sem{1}(m_2)$: the
    only possible non-zero term in this infinite sum is for
    $m_2 = \eps$ and $m_1 = \lambda(a)$, which is only possible if
    $m=\lambda(a)$. Thus the infinite sum exists by
    Lemma~\ref{lem.zsum}: it is equal to $\pi(a)$ if $\lambda(a)=m$,
    and $0$ otherwise. On the other hand, the family
    $\{\pi(w)\indic{a}(w)\mid w\in A^*,\lambda(w)= m\}$ contains at
    most one non-zero element, $\pi(a)$, and its sum therefore exists,
    equal to $\pi(a)$ if $\lambda(a)=m$, 0 otherwise.

  \item If $E$ is \emph{true} or \emph{false}, the equivalence is
    trivial.
    
  \item Suppose that $E=F+G$ and that the induction hypothesis holds
    for $F$ and $G$. 
%
    If~$E_{\lambda,\pi}$ is valid, in particular, it is also the case
    for all its subexpressions, and we can apply for them the
    induction hypothesis. Let $m\in M$. We know that
    $\sem{E_{\lambda,\pi}}(m)$ is defined, and then
    \begin{align*}
      \sem{E_{\lambda,\pi}}(m)
      &=\sem{F_{\lambda,\pi}}(m)+\sem{G_{\lambda,\pi}}(m)
      &\text{(by definition)}\\
      &=\sum_{\lambda(w)= m}\pi(w)\indic{F}(w)
        +\sum_{\lambda(w)= m}\pi(w)\indic{G}(w)
      &\text{(by induction)}\\
      &=\sum_{\lambda(w)= m}
        (\pi(w)\indic{F}(w)+\pi(w)\indic{G}(w))
      &\text{(by Lemma~\ref{lem.linsum})}\\
      &=\sum_{\lambda(w)= m}\pi(w)\indic{F+G}(w)
      &\text{(by unambiguity)}
    \end{align*}

    Reciprocally, assume that
    $\sum_{\lambda(w)= m}\pi(w)\indic{E'}(w)$ exists, for all $m\in M$
    and subexpressions $E'$ of~$E$. Let $m\in M$. In particular,
    $\sum_{\lambda(w)= m}\pi(w)\indic{F}(w)$ and
    $\sum_{\lambda(w)= m}\pi(w)\indic{G}(w)$ exist and are equal to
    $\sem{F_{\lambda,\pi}}(m)$ and $\sem{G_{\lambda,\pi}}(m)$,
    respectively. Then,
    \begin{align*}
      \sum_{\lambda(w)= m}\pi(w)\indic{F}(w)
      +\sum_{\lambda(w)= m}\pi(w)\indic{G}(w)
      &=\sem{F_{\lambda,\pi}}(m)+\sem{G_{\lambda,\pi}}(m)=\sem{E_{\lambda,\pi}}(m)
    \end{align*}
    On the other hand, using Lemma~\ref{lem.linsum}, and relying on
    the unambiguity of $F+G$, 
    \begin{align*}
      \sum_{\lambda(w)= m}\pi(w)\indic{F}(w)
      +\sum_{\lambda(w)= m}\pi(w)\indic{G}(w)
      &=\sum_{\lambda(w)= m}
        (\pi(w)\indic{F}(w)+\pi(w)\indic{G}(w))\\
      &=\sum_{\lambda(w)= m}\pi(w)\indic{F+G}(w)
    \end{align*}
    As a conclusion, $\sem{E_{\lambda,\pi}}(m)$ is defined and is equal
    to $\sum_{\lambda(w)= m}\pi(w)\indic{E}(w)$.
    
  \item Suppose that $E=F\cdot G$ and that the induction hypothesis
    holds for $F$ and $G$. For $m\in M$, we let
    $\psplit{m}=\{(u,v)\in (A^*)^2\mid \exists m_1,m_2\in M \;
    m_1m_2=m, \lambda(u)=m_1,\lambda(v)=m_2\}$.  Remark that it can
    also be written
    $\{(u,v)\mid \exists w\in A^*\; \lambda(w)=m, w=uv\}$, because
    $\lambda$ is a morphism.

    If $E_{\lambda,\pi}$ is valid, in particular, it is also the case
    for all its subexpressions, and we can apply for them the
    induction hypothesis. Let $m\in M$. The value
    $\sem{E_{\lambda,\pi}}(m)$ is defined and is equal to
    \begin{align*}
      &\sum_{m_1m_2=m}
        \sem{\efa{F}}(m_1)\times \sem{\efa{G}}(m_2)
      &\text{(by definition)}\\
      &=\sum_{m_1m_2=m}\Big(
        \sum_{\lambda(u)=m_1}\wautac(u)\indic{F}(u)\Big)
        \times \Big(
        \sum_{\lambda(v)=m_2}\wautac(v)\indic{G}(v)\Big)
      &\text{(by induction)}\\
      &=\sum_{m_1m_2=m}
        \sum_{\lambda(u)=m_1}
        \sum_{\lambda(v)=m_2}
        \wautac(u)\indic{F}(u)\wautac(v)\indic{G}(v)
      &\text{(by~\axref{ax3} twice)}\\
      &=\sum_{m_1m_2=m}
        \sum_{\lambda(u)=m_1, \lambda(v)=m_2}
        \wautac(u)\indic{F}(u)\wautac(v)\indic{G}(v)
      &\text{(by~\axref{ax4})}\\
      &=\sum_{(u,v)\in\psplit{m}}
        \wautac(u)\indic{F}(u)\wautac(v)\indic{G}(v)
      &\text{(by~\axref{ax4})}\\
      &=\sum_{\lambda(w)= m}
        \sum_{uv=w}
        \wautac(u)\indic{F}(u)\wautac(v)\indic{G}(v)
      &\makebox[0cm][r]{\text{(by \axref{ax5} and finite sum)}}\\
      &=\sum_{\lambda(w)= m}
        \sum_{uv=w}
        \wautac(u)\wautac(v)\indic{F}(u)\indic{G}(v)\\
      &=\sum_{\lambda(w)= m}
        \sum_{uv=w}
        \wautac(w)\indic{F}(u)\indic{G}(v)
      &\text{($\pi$ is a morphism)}\\
      &=\sum_{\lambda(w)= m}\wautac(w)
        \sum_{uv=w}
        \indic{F}(u)\indic{G}(v)
      &\makebox[0cm][r]{\text{(distributivity over finite sums)}}\\
      &=\sum_{\lambda(w)= m}\wautac(w)\indic{F\cdot G}(w)
      &\text{(by unambiguity)}
    \end{align*}

    Reciprocally, assume now that
    $\sum_{\lambda(w)= m}\pi(w)\indic{E'}(w)$ exists for all $m\in M$
    and all subexpressions $E'$ of~$E$. Let $m, m_1, m_2\in M$. In
    particular, $\sum_{\lambda(u)= m_1}\pi(u)\indic{F}(u)$, and
    $\sum_{\lambda(v)= m_2}\pi(v)\indic{G}(v)$ exist. So does the
    product of the two latter sums, that is moreover equal to
    \begin{align*}
      &\sum_{\lambda(u)=m_1}
        \sum_{\lambda(v)=m_2}
        \wautac(u)\indic{F}(u)\wautac(v)\indic{G}(v)=\sum_{\lambda(u)=m_1,
        \lambda(v)=m_2} 
        \wautac(u)\indic{F}(u)\wautac(v)\indic{G}(v)
    \end{align*}
    by the application of \axref{ax3} and \axref{ax4}. 
    Then, the sum
    $\sum_{\lambda(w)= m}\pi(w)\indic{F\cdot G}(w)$ is equal to
    \begin{align*}
      &\sum_{\lambda(w)= m}\wautac(w)
        \sum_{uv=w}
      \indic{F}(u)\indic{G}(v)
      &\text{(by unambiguity)}\\
      &=\sum_{\lambda(w)= m}
        \sum_{uv=w}
      \wautac(w)\indic{F}(u)\indic{G}(v)
      &\makebox[0cm][r]{\text{(distributivity over finite sums)}}\\
      &=\sum_{\lambda(w)= m}
        \sum_{uv=w}
      \wautac(u)\wautac(v)\indic{F}(u)\indic{G}(v)
      &\text{($\pi$ is a morphism)}\\
      &=\sum_{(u,v)\in\psplit{m}}
        \wautac(u)\indic{F}(u)\wautac(v)\indic{G}(v)
      &\text{(by \axref{ax4})}\\
      &=\sum_{m_1m_2=m}
      \sum_{\lambda(u)=m_1, \lambda(v)=m_2}
      \wautac(u)\indic{F}(u)\wautac(v)\indic{G}(v)
      &\text{(by~\axref{ax5})}\\
      &=\sum_{m_1m_2=m}\Big(
      \sum_{\lambda(u)=m_1}\wautac(u)\indic{F}(u)\Big)\times
      \Big(
      \sum_{\lambda(v)=m_2}\wautac(v)\indic{G}(v)\Big)\\
      &=\sum_{m_1m_2=m}
      \sem{\efa{F}}(m_1)\times \sem{\efa{G}}(m_2)
      &\text{(by induction)}\\
      &=\sem{(F\cdot G)_{\lambda,\pi}}(m)
      &\text{(by definition)}
    \end{align*}
    As a conclusion $\sem{E_{\lambda,\pi}}(m)$ exists and is equal to
    $\sum_{\lambda(w)= m}\pi(w)\indic{E}(w)$.

  \item Suppose that $E=F^*$ and that the induction hypothesis holds
    for $F$. By an easy induction using the previous case, the
    equivalence holds for $F^n $ for all $n\geq 0$:
    $(F_{\lambda,\pi})^n$ is valid if and only if
    $\sum_{\lambda(w)= m }\pi(w)\indic{F^n}(w)$ is defined for all
    $m\in M$, in which case
    $\sem{(F_{\lambda,\pi})^n}(m)=\sum_{\lambda(w)= m
    }\pi(w)\indic{F^n}(w)$.

    If $E_{\lambda,\pi}$ is valid, then for all $m\in M$,
    \begin{align*}
      \sem{E_{\lambda,\pi}}(m)
      &= \sem{(F_{\lambda,\pi})^*}(m)\\
      &=\sum_{n=0}^{\infty}\sem{(\efa{F})^n}(m)
      &\text{(by definition)}\\
      &=\sum_{n=0}^{\infty}\sum_{\lambda(w)= m }\pi(w)\indic{F^n}(w)\\
      &=\sum_{\lambda(w)= m}
        \sum_{n=0}^{\infty}
        \wautac(w)\indic{F^n}(w)
      &\text{(by Lemma~\ref{lem.permsum} and unambiguity)}\\
      &=\sum_{\lambda(w)= m}\wautac(w)
        \sum_{n=0}^{\infty}
        \indic{F^n}(w)
      &\text{(by~\axref{ax3})}\\
      &=\sum_{\lambda(w)= m}\wautac(w)
        \indic{F^*}(w)
      &\text{(by unambiguity)}
    \end{align*}

    Reciprocally, assume that
    $\sum_{\lambda(w)= m}\pi(w)\indic{E'}(w)$ exists for all $m\in M$
    and all subexpressions $E'$ of $E$. Remark that since $E$ is
    unambiguous, $\sum_{n=0}^{\infty}\indic{F^n}(w)$ exists for all
    $w$. Then, for all $m\in M$, 
    \begin{align*}
      \sum_{\lambda(w)= m}\pi(w)\indic{F^*}(w)
      &=\sum_{\lambda(w)= m}\wautac(w)
        \sum_{n=0}^{\infty}
        \indic{F^n}(w)
      &\text{(by unambiguity)}\\
      &=\sum_{\lambda(w)= m}
        \sum_{n=0}^{\infty}
        \wautac(w)\indic{F^n}(w)
      &\text{(by~\axref{ax3})}\\
      &=\sum_{n=0}^{\infty}\sum_{\lambda(w)= m }\pi(w)\indic{F^n}(w)
      &\text{(by Lemma~\ref{lem.permsum})}\\
      &=\sum_{n=0}^{\infty}\sem{(\efa{F})^n}(m)\\
      &=\sem{E_{\lambda,\pi}}(m)
      &\text{(by definition)}\tag*{\qedhere}
    \end{align*}
  \end{itemize}
 \end{proof}


Starting from a weighted expression $W$, and in order to use
Lemma~\ref{lem.central} which only applies to unambiguous Kleene expressions,
we will modify $W$ to interpret it as an unambiguous Kleene expression.
We define its \emph{indexed expression} $\col W$ as the Kleene expression
over an alphabet being a finite subset of $(\K\cup M)\times \N$,
obtained by replacing each of its atomic subexpression $\ell \in \K\cup M$
by a letter $(\ell, i)\in (\K\cup M)\times \N$ where $i$ is a unique index
(starting from $0$ for the leftmost one) associated with each
atomic subexpression. For instance, with the weighted expression
$W=(2\cdot a + 3\cdot b)^* \cdot (a + 5\cdot b + 3)$, one associates the
indexed expression
$\col W = ((2,0) \cdot (a,1) + (3,2) \cdot (b,3))^* \cdot ((a,4) +
(5,5)\cdot (b,6) + (3,7))$. From the indexed expression, one can
recover the initial expression, by forgetting about the
index. Formally, we let $\lambda$ be the morphism from
$((\K\cup M)\times \N)^*$ to $M$ such that $\lambda(x,n)=x$ if
$x\in M$ and $\eps$ otherwise, and $\pi$ be the morphism from
$((\K\cup M)\times \N)^*$ to~$\K$ such that $\pi(x,n)=x$ if $x\in\K$
and $1$ otherwise. For the above example, 
$\col W_{\lambda,\pi} = ((\eps \cdot 2)\cdot (a\cdot 1) + (\eps\cdot
3)\cdot (b\cdot 1))^* \cdot ((a\cdot 1) + (\eps\cdot 5)\cdot (b\cdot
1) + (\eps \cdot 3))$, which is equivalent to $W$. More generally, we
obtain:
\begin{lemma}\label{lem.lpequiv}
  For all weighted expressions $W$ over $M$, $\col{W}_{\lambda,\pi}$
  is valid if and only if $W$ is valid. When valid, they have the same
  semantics.
\end{lemma}
\begin{proof}

  We proceed by induction.
  \begin{itemize}
  \item If $W=k\in\K$, we have $\col{W}=(k,0)$ and
    $\col{W}_{\lambda,\pi}=\lambda(k,0)\pi(k,0)=\eps\cdot k$, so that
    the result is trivial.
  \item If $W=m\in M$, we have $\col{W}=(m,0)$ and
    $\col{W}_{\lambda,\pi}=\lambda(m,0)\pi(m,0)=m\cdot 1$, which
    also allows us to conclude immediately.
  \item If $W=U+V$, then we see easily that
    $\col{W}_{\lambda,\pi}=\col{U}_{\lambda,\pi}+\col{V}_{\lambda,\pi}$
    that is valid if and only $\col{U}_{\lambda,\pi}$ and
    $\col{V}_{\lambda,\pi}$ are, that is, by induction
    hypothesis, if and only if $U$ and $V$ are valid. In this case,
    the equivalence of $W$ and $\col{W}_{\lambda,\pi}$ holds.
  \item If $W=U\cdot V$, then we have
    $\col{W}_{\lambda,\pi}=\col{U}_{\lambda,\pi}\cdot\col{V}_{\lambda,\pi}$ and
    we conclude similarly. 

    
  \item If $W=U^*$, we have again
    $\col{W}_{\lambda,\pi}=(\col{U}_{\lambda,\pi})^*$ and the result
    follows as before.
    %
  \end{itemize}
\end{proof}

We would like to conclude by combining this result with
Lemma~\ref{lem.central} and by using the pre-rationality of the monoid
$M$, as in Theorem~\ref{theo.automata}. However, $\col W$ might not be
unambiguous as expected, as shown by the example $W=(m^*)^*$, with
$m\in M$, that gives rise to the (ambiguous) Kleene expression
$\col W = (((m,0))^*)^*$: indeed, the word $(m,0)(m,0)$ has several
possible decompositions in the semantics of $\col W$. To patch this
last issue, we simply incorporate a dummy marker after each Kleene
star as follows: from a weighted expression~$W$, $\una{W}$ is
inductively defined by:
\begin{itemize}
\item if $W$ is an atomic expression, $\una{W}=W$;
\item if $W=U+V$ then $\una{W}=\una{U}+\una{V}$;
\item if $W=U\cdot V$ then $\una{W}=\una{U}\cdot \una{V}$;
\item if $W=U^*$ then $\una{W}=(\una{U})^*\cdot 1$, with $1$ being the
  neutral element of the semiring $\K$. 
\end{itemize}
We directly obtain:
\begin{lemma}\label{lem.unambiguity}
  Let $W$ be a weighted expression. The Kleene expression
  $\col{\una{W}}$ is \mbox{unambiguous}.
\end{lemma}
\begin{proof}
  The proof goes by an easy induction since letters appearing in all
  subexpressions of $W$ are different. For the Kleene star $W^*$, we
  also rely on the fact that
  $\col{\una {W^*}} = (\col {\una W} )^*\cdot (1,n)$, with $n\in \N$,
  so that every word of $\Lang(\col{\una {W^*}})$ can only be split in
  a unique way because of the index $n$.
\end{proof}

We are now ready to conclude the proof of Theorem~\ref{thm.expok},
moreover showing that for all weighted expressions $W$ and $m\in M$,
$\sem{W}(m)=
\sum_{\lambda(w)=m}\pi(w)\indic{\col{\una{W}}}(w)$. Indeed, operation
$\una{\cdot}$ does not change the semantics of an expression, and
therefore, $\una{W}$ is valid if and only if $W$ is valid, in which
case they share the same semantics. Using the result of
Lemma~\ref{lem.unambiguity}, we can apply Lemma~\ref{lem.lpequiv}: $W$
is valid if and only if $\col{\una{W}}_{\lambda,\pi}$ is valid, in which case
they are equivalent.  Let $L=\lang{\col{\una{W}}}\cap \invlab{}(m)$.
Since $M$ is pre-rational, $L$ is a rational language, and
$\sum_{w\in L}\pi(w)$ exists. Moreover,
\begin{align*}
  \sum_{w\in L}\pi(w)
  &=\sum_{\lambda(w)=m}\pi(w)\indic{\col{\una{W}}}(w)\\
  &=\sem{\col{\una{W}}_{\lambda,\pi}}(m)
  &\text{(by Lemma~\ref{lem.central})}\\
  &=\sem{\una{W}}(m)
  &\text{(by Lemma~\ref{lem.lpequiv})}\\  
  &=\sem{W}(m)
  &\text{($W$ and $\una W$ are equivalent).}
\end{align*}

\section{A Kleene-Like Theorem}\label{sec.kth}
Our main result 
is the following Kleene-like theorem,
stating the constructive equivalence between
expressions and automata
over a pre-rational monoid and 
weighted over a rationally additive semiring.

\begin{theorem}\label{theo.eqatoe}
  Let $\K$ be a rationally additive semiring,
  and $M$ be a pre-rational monoid. Let $s\in\series$ be a
  series. Then $s$ is the semantics of some $\K$-automaton over $M$ if
  and only if it is the semantics of some $\K$-expression over
  $M$.
\end{theorem}

The rest of this section is devoted to the proof of this theorem, that
consists in constructive translations of automata into equivalent
expressions, and vice versa.

\smallskip
\noindent\textbf{From Automata to Expressions.}
The idea is to build a $\K$-expression from an unambiguous expression
generating the accepting runs of the automaton. Let
$\Ac=(Q,\Delta,I,F)$ be a $\K$-automaton over $M$. By applying the
result of \cite{BooEve71}, we build an unambiguous Kleene expression
$E$ over $\Delta^*$ recognising the language~$R_\Ac$ of the accepting runs
of $\Ac$.
By Lemma~\ref{lem.central}, that we can apply on $E$ since
$E_{\lambda_\Ac,\pi_\Ac}$ is valid (by Theorem~\ref{thm.expok}), we have
\begin{align*}
  \evalm{\efa{E}}
  &= \sum_{\lambda_\Ac(w)=m}\pi_\Ac(w)\indic{E}(w) =\sum\limits_{\substack{w\in R_\Ac\mid \lambda_\Ac(w)= m}}\pi_\Ac(w)=\evalm{\Ac}.
\end{align*}
the second equality coming from \eqref{eq:caracteristic}, since
$\Lang(E) = R_\Ac$.

\smallskip
\noindent\textbf{From Expressions to Automata.}
We have shown in the previous section how, from a $\K$-expression $E$
over $M$, we can construct an unambiguous Kleene expression
$\col{\una{E}}$ and two morphisms $\lambda$ and $\pi$ from\footnote{
  As before, in fact, we work with a finite subset of this set.  }
$(\K\cup M)\times\N$ to respectively $M$ and $\K$, such that
$\col{\una{E}}_{\lambda,\pi}$ is equivalent to $E$, and by
Theorem~\ref{thm.expok},
$\evalm{E}=\sum_{\lambda(w)=m}\pi(w)\indic{\col{\una{E}}}(w)$. We let
$\{0,\ldots,n\}$ be the set of indices used in $\col{\una{E}}$.

By \cite{BooEve71}, we can build (for instance, by considering the
position automaton) from $\col{\una{E}}$ an equivalent unambiguous
Boolean automaton $\Ac=(Q,\Delta,I,F)$ with
$\Delta \subseteq Q\times \big((\K\cup M)\times\N\big) \times Q$ its
set of transitions labelled by indexed atomic elements appearing in
$E$. Here, unambiguous means as usual that every accepted word in
$\Ac$ is associated with a unique accepting run.

From $\Ac$, we build a $\K$-automaton
$\Bc = (Q\times \{0,\ldots,n\}, \Delta', I\times \{0\}, F\times
\{0,\ldots,n\})$ over $M$ with transitions defined as follows: for all
transitions $(p,(m,i),q)\in \Delta$, with $m\in M$, we add the
transition $((p,j), m, 1, (q,i))\in \Delta'$, and for all transitions
$(p,(k,i),q)\in \Delta$, with $k\in \K$, we add the transition
$((p,j), \eps, k, (q,i))\in \Delta'$. The transfer of indices from
letters to states enables us to obtain a bijection $f$ from accepted
words of $\Ac$ to accepting runs of $\Bc$. Moreover, this bijection
preserves the labels and weights, meaning that for all
$u = (x_0,i_0)\cdots(x_m,i_m)$ accepted by $\Ac$, we have
$\lambda(u)=\lambda_\Bc(f(u))$, and $\pi(u)=\pi_\Bc(f(u))$.
Therefore, by applying the change of variable $w=f(u)$, we obtain
\[\evalm{\Bc}=\smashoperator[r]{\sum_{w\in R_\Bc\cap\lambda_{\Bc}^{-1}(m)}}\quad\pi_\Bc(w) = 
  \smashoperator[r]{\sum_{u\in \Lang(\col{\una{E}})\cap\lambda^{-1}(m)}}\qquad\pi(u)
  =\sum_{\lambda(u)=m}\pi(u)\indic{\col{\una{E}}}(u)= \evalm{E}.\]

\section{Dealing with Ambiguity}\label{sec:unambiguous}

We have already encountered ambiguity in the context of the Boolean
semiring and free monoids. We now study this notion for weighted
expressions and automata.
To do so, we use the rationally additive
semiring $(\ninf= \N\cup\set{\infty},+,\times, 0, 1)$ where all
infinite sums exist: in particular, the sum over a family containing
an infinite number of non-zero values is $\infty$, and otherwise the
sum is equal to the finite sum over the support of the family. We call
this semiring the \emph{counting semiring}.

\begin{definition}
  Given a $\K$-expression $W$ over the monoid $M$, the \emph{ambiguity
    $\amb W m$ of $W$ at $m$} is a value in $\ninf$ defined
  inductively over $W$ as follows:
  \begin{itemize}
  \item for $W=n\in M$, $\amb n m = 1$ if $n=m$, and $0$ otherwise;
  \item for $W=k\in \K$, $\amb k m = 1$ if $m=\varepsilon_M$, and $0$
    otherwise;
  \item for $W=U+V$, $\amb{U+V}m = \amb U m + \amb V m$;
  \item
    for $W=U\cdot V$, $\amb{U\cdot V} m = \sum_{m_1m_2=m}\amb U{m_1}\times \amb V{m_2}$;
  \item for $W=U^*$, $\amb{U^*} m = \sum_{n\in \N} \amb{U^n}m$.
  \end{itemize}
  An expression is called \emph{unambiguous} if its ambiguity at every point is at most 1.
\end{definition}
For instance, the expression $W = 2\cdot a + 3\cdot a\cdot a$ over the
free monoid $\{a\}^*$ is unambiguous, while $W^*$ has ambiguity $2$ at
the word $aaa = a\cdot aa = aa\cdot a$.

The attentive reader may have noticed that the ambiguity of $W$ is
exactly the semantics of $W$ where every atomic weight of $\K$ is
replaced with the unit $1$ of $\ninf$. Given two rationally additive
semirings $\K_1$ and $\K_2$, $\K_1\times \K_2$ is also a rationally
additive semiring with the natural component-wise operations. In
particular, given a $\K$-expression $W$, we can define a
$\K\times \ninf$-expression $W'$ by replacing every weight $k\in \K$
appearing in $W$ by $(k,1)\in \K\times \ninf$. Then, the ambiguity of
$W$ at $m$ is the second component of the weight $\sem{W'}(m)$.

\begin{definition}
  Given a $\K$-automaton $\Ac$ over the monoid $M$, the
  \emph{ambiguity of $\Ac$ at $m$} is a value in $\ninf$ defined as
  the number (potentially $\infty$) of runs with label $m$.  An
  automaton is called \emph{unambiguous} if its ambiguity at every point
  is at most $1$.
\end{definition}
Just as for expressions, the ambiguity of an automaton may be viewed
as the semantics of the automaton where the weights of transitions are
replaced by the unit of $\ninf$.  Given $\Ac$ over $\K$, we can define
$\Ac'$ by replacing all weights $k\in \K$ of transitions by
$(k,1)\in \K\times \ninf$. Then the ambiguity of $\Ac$ at $m$ is
exactly the second component of $\sem {\Ac'}(m)$.
Now we claim:
\begin{theorem}\label{theo.nonamb}
  Let $\K$ be a rationally additive semiring, $M$ be a pre-rational
  monoid, $s\in \series$, and $k\in\N$.  Then, $s$ is the semantics of
  a $\K$-automaton over $M$ of ambiguity $k$ if and only if it is the
  semantics of a $\K$-expression over $M$ of ambiguity $k$.
\end{theorem}
\begin{proof}
  The procedures of section~\ref{sec.kth}
  to go from expressions to automata and back, over a
  pre-rational monoid $M$, preserve ambiguity. Indeed,
  the two constructions used to prove Theorem~\ref{theo.eqatoe} do not
  introduce new weights. Thus, starting from a $\K$-expression $W$,
  one considers the $\K\times\ninf$-expression $W'$ defined
  above. Transforming $W'$ into an automaton preserves the semantics,
  and all the transitions have a second component equal to $1$. Thus,
  the second component of the semantics, which is preserved, is
  exactly the ambiguity of the automaton. Forgetting about the second
  component, we get the result. Note that converting $W$ to $W'$ is
  not actually a necessary step to build the automaton, it is simply a
  mental crutch to make the argument simpler. Symmetrically when
  going from automata to expressions, the transformation does not
  introduce new weights and thus preserves ambiguity.
\end{proof}

\section{Free Inverse Monoids and Applications to Walking Automata}
\label{sec:fim}

We conclude this article by demonstrating why our model is able to
encompass and reason about the usual models of two-way automata and
tree-walking automata. To do so, we consider the free inverse monoid,
as it was observed by Pécuchet
\cite{Pecu85} to be linked with this model.
Dicky and Janin even gave in \cite[Theorem 3.21]{DJ15}
the equivalence in the boolean case between two-way automata
and regular expressions, using this monoid.

Let $A$ be a finite alphabet, and $\ovl{A}=\{\ovl{a}\mid a\in A\}$ be
a copy of $A$. We define the function~$\dagger\colon(A\cup\ovl{A})^*\to (A\cup\ovl{A})^*$ inductively as:
$  \mydag{\epsilon}=\epsilon$, 
  $\mydag{(ua)}=\ovl{a}\mydag{u}$,  and
  $\mydag{(u\ovl{a})}=a\mydag{u}
$.

\begin{definition}
  The \emph{free inverse monoid} ${\fim}(A)$ generated by a finite
  alphabet~$A$ is the quotient of~$(A\cup\ovl{A})^*$ by the following
  equivalence relations:
  \begin{itemize}
  \item ``$\mydag{x}$ and $x$ are pseudo-inverses'': for all
    $x\in(A\cup\ovl{A})^*$, $x\mydag{x}x=x$, and
    $\mydag{x}x\mydag{x}=\mydag{x}$;
  \item ``idempotent elements commute'': for
  all $x,y\in(A\cup\ovl{A})^*$:
  $x\mydag{x}y\mydag{y}=y\mydag{y}x\mydag{x}$.
  \end{itemize} 
\end{definition}

Notice that $x\mydag{x}$ are indeed idempotent elements of the free
inverse monoid, since
$(x\mydag{x})(x\mydag{x})=(x\mydag{x}x)\mydag{x}=x
\mydag{x}$.

The elements of this monoid are convenientely represented via
tree-like structures, the Munn bi-rooted trees \cite{M74}.
They are directed graphs, whose underlying undirected graph
is a tree, and two special nodes are marked,
the \emph{initial} and the \emph{final} one.
Examples of elements of the monoid with
their Munn tree representation
are given in Figure~\ref{fig.morph}.
Note that if you see $a\in A$ as the traversal of
an edge labelled by $a$, and $\ovl{a}$ its traversal in reverse,
an element of $(A\cup\ovl{A})^*$ describes a complete
walk over the graph of the corresponding element of $\fim(A)$.

\tikzstyle{position} = [state,draw=none,minimum size=0pt,inner sep=1pt]

\begin{figure}[tbp]
  \centering
  \scalebox{.9}{
  \begin{tikzpicture}[yscale=.8]
    \node[position, initial above](0){};
    \node[position, accepting below,below left of=0,yshift=5mm](1){};
    
    \draw (0) edge node[above left]{$\ell$} (1);

    \begin{scope}[xshift=1cm] 
      \node[position, accepting below](0){};
      \node[position, initial above,below right of=0,yshift=5mm](1){};
      
      \draw (0) edge node[above right]{$r$} (1);
    \end{scope}

    \begin{scope}[xshift=4.5cm] 
      \node[position](1){};
      \node[position, initial above,below left of=1,yshift=5mm](0){};
      \node[position, below right of =1,yshift=5mm](2){};
      \node[position, above right of =2,yshift=-5mm](3){};
      \node[position, accepting below, below right of =3,yshift=5mm](4){};
      \path (1) edge node[above left]{$\ell$} (0)
      (1) edge node[above right]{$r$} (2)
      (3) edge node[above left]{$\ell$} (2)
      (3) edge node[above right]{$r$} (4);
    \end{scope}

    \begin{scope}[xshift=10cm]
      \node[position, initial above](0) {};
      \node[position, below left of=0,yshift=5mm](1) {};
      \node[position, below right of=0,yshift=5mm](2) {};
      \node[position, below left of=2,yshift=5mm](3) {};
      \node[position, below right of=2,yshift=5mm](4) {};
      \node[position, below left of=4 ,accepting below,yshift=5mm](5) {};

      \path (0) edge node[above left]{$\ell$} (1)
      (0) edge node[above right]{$r$} (2)
      (2) edge node[above left]{$\ell$} (3)
      (2) edge node[above right]{$r$} (4) 
      (4) edge node[above left]{$\ell$} (5);
    \end{scope}
  \end{tikzpicture}
  }
  \caption{Munn bi-rooted trees of the elements of $\fim(\{\ell,r\})$: $\ell$,
    $\bar r$, $\bar \ell r \bar \ell r$, and $(\ell \bar \ell r)^2 \ell$.}
  \label{fig.morph}%
\end{figure}
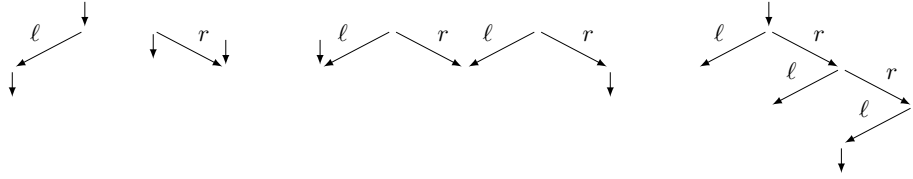

With this tree representation in mind, we see that every element of
$\fim(A)$ has finitely many prefixes,
since such a prefix is a subtree of $x$, with the same initial node.
Thanks to Lemma~\ref{lem.pref}, we obtain
\begin{proposition}
  The free inverse monoid is pre-rational.
\end{proposition}

We can thus apply our results on this pre-rational monoid, for
instance by considering expressions.  In the Boolean semiring, for
example, the expression $(\ell\cdot \bar{\ell}\cdot r)^*\cdot\ell$
describes the language of Munn bi-rooted trees that are
``right-combs'' (see the rightmost tree of Figure~\ref{fig.morph}),
when considering~$\ell$ to be left children, and~$r$ right ones. The
initial node is at the top while the final one is the farthest away
from it. We can add weights to this expression: in the tropical
semiring $(\Z\cup\set{{-}\infty,{+}\infty},\sup,+,{-}\infty, 0)$, the
unambiguous expression
$(\ell\cdot \bar{\ell}\cdot r \cdot 1)^*\cdot \ell$ associates with a
comb the length of its rightmost branch. More generally, the
expression
$W=\big[\sum_{a\in A} \big(a \cdot 1 + \bar a \cdot (-1)\big)\big]^*$
computes the (signed) length of the path linking the initial and final
nodes in any Munn bi-rooted tree over alphabet $A$: each tree is
associated with the difference between the number of positive letters of
$A$ and the number of negative letters of $\bar A$ of the unique
acyclic path linking the initial node to the final node.
On the trees of Figure~\ref{fig.morph},
these lengths are respectively $1$, $-1$, $0$, $3$.
They represent the difference of ``levels'' in-between the initial and
final nodes. Each tree is associated with many decompositions in the
semantics of the expression $W$, but all of them have the same weight
(and the chosen semiring has an idempotent sum operation).

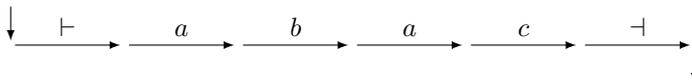
\begin{figure}[b]
  \begin{tikzpicture}[->,>=latex]
    \node[position,initial above] (0) {};
    \node[position,right of=0] (1) {};
    \node[position,right of=1] (2) {};
    \node[position,right of=2] (3) {};
    \node[position,right of=3] (4) {};
    \node[position,right of=4] (5) {};
    \node[position,right of=5,accepting below] (6) {};
    
    \draw (0) edge node[above]{$\bmark$} (1)
    (1) edge node[above]{$a$} (2)
    (2) edge node[above]{$b$} (3)
    (3) edge node[above]{$a$} (4)
    (4) edge node[above]{$c$} (5)
    (5) edge node[above]{$\emark$} (6);
  \end{tikzpicture}
  \caption{Munn bi-rooted tree of the ``word'' $\bmark abac \emark$.}
  \label{fig.word}
\end{figure}

\smallskip\noindent
\textbf{Two-way Automata.}  Over an alphabet~$A$, we can consider the
free inverse monoid $\fim(A\uplus\{\bmark,\emark\})$, with two fresh
symbols $\bmark$ and $\emark$ that will help us distinguish the
leftmost and rightmost letters of the word. To model two-wayness, only
certain elements of $\fim(A\uplus\{\bmark,\emark\})$ are of interest,
namely elements of $\bmark A^* \emark$, that have linear Munn
bi-rooted trees with the initial node at the leftmost position, and
the final node at the rightmost one. The Munn bi-rooted tree
representation of such an element is given in Figure~\ref{fig.word}.

We thus consider weighted automata and expressions over $\fim(A)$ with
weights in $\K$, a rationally additive semiring, and restrict our
attention to words of $\bmark A^* \emark$.
%
From an automata perspective, this is a way to define the usual model
of two-way automata, a forward movement of a two-way automaton being
simulated by reading of a letter in $A$ while a backward movement
is simulated by reading a letter in $\bar A$.
Indeed, our model of weighted automata over $\fim(A)$ can also be
simulated by the usual two-way weighted automata, since non-atomic
elements of the monoid can be split into atomic elements. 
Therefore, in this specific context, Theorem~\ref{theo.eqatoe} gives a
new way to express the semantics of two-way weighted automata (over a
rationally additive semiring) by using expressions.

Consider for example the function that maps a word $\bmark w\emark$
with $w=w_0\cdots w_{n-1}\in \{a,b\}^*$ to the set of words
$\{(w_{n-1}\cdots w_0)^k\mid k\in \N\}$. Considering the semiring of
regular languages, a weighted expression describing this function is
  \[\left(\bmark\cdot (a+b)^*\cdot\emark\cdot
  \ovl{\emark}\cdot(\ovl{a}\cdot\{a\}+\ovl{b}\cdot\{b\})^*\cdot \ovl{\bmark}\right)^*
\cdot\bmark\cdot(a+b)^*\cdot\emark.\]
Notice the last pass over the word that allows one to finish the
reading on the rightmost position, i.e.~the final node. 

Consider the alphabet $A = \{\mathbf 0, \mathbf 1\}$. For a word
$w\in A^*$, let $w_{|2}$ denote the rational number between 0 and 1
that is written as $0.w$ in binary. Then, consider the following weighted expression
with weights in $(\Q_+\cup\{+\infty\},+,\times,0,1)$:
\[W = \bmark \cdot \Big(\mathbf 0 \cdot \frac 1 2 + \mathbf 1 \cdot
\frac 1 2\Big)^* \cdot \mathbf 1 \cdot \frac 1 2 \cdot (\mathbf 0 +
\mathbf 1)^* \cdot \emark. \]
It associates with a word $\bmark w \emark$ the value $w_{|2}$, since it non-deterministically
chooses a position~$i$ labelled by $\mathbf 1$ in $w$ and computes the value $1/2^i$.
By considering the expression \[(W \cdot \ovl \emark \cdot(\ovl{\mathbf 0} +
\ovl{\mathbf 1})^*\cdot\ovl\bmark)^* \cdot W.\]
that consists in repeating the computation
of $W$ any number of times (at least once), with a reset of the word
in-between, we associate with a word $\bmark w \emark$ the value
$\sum_{n=1}^\infty w_{|2}^n = w_{|2}/(1-w_{|2})$. 

\smallskip\noindent \textbf{Tree-Walking Automata.}  Another model
captured by our approach is the one of tree-walking automata.
These are automata whose head moves on the nodes of a rooted tree
of a bounded arity $m$.
As for words before, we can encode such trees labelled with a
finite alphabet~$A$ by elements of $\fim(A')$ with an extended
alphabet $A' = (\{0,\ldots,m-1\}\cup\{\top\})\times A\cup\{\bot\}$.
In~elements of $\fim(A')$, nodes contain no information, only edges
do. The idea is thus to simulate the root of a tree labelled with $a$
by a single node labelled with $(\top,a)$; the $i$-th child of a node,
labelled with $a\in A$, will be simulated with a node of label
$(i,a)$; finally, under each leaf of the tree, we add a node labelled
with $\bot$. The root of the tree will be both the initial and the
final node of the encoding, simulating a tradition of tree-walking
automata to start and end in the root of the tree (without loss of
generality). 

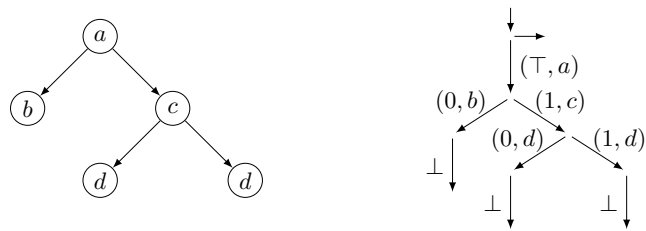
\begin{figure}[tbp]
  \centering\scalebox{.9}{\begin{tikzpicture}
    \begin{scope}[state/.append style={,inner sep=2pt,minimum size=5mm}]
      \node[state] (a) {$a$}; \node[state] (b) [below left of = a]
      {$b$};
      \node[state] (c) [below right of = a] {$c$}; \node[state] (d)
      [below left of = c] {$d$}; \node[state] (e) [below right of = c]
      {$d$};

      \draw (a) -> (b); \draw (a)->(c); \draw (c)->(d); \draw
      (c)->(e); 
    \end{scope}
    
    \begin{scope}[xshift = 6cm, node distance=1.2cm,yscale=.8]
      \node[position,initial above, accepting right] (t) {};
      \node[position] (p) [below of = t,yshift=3mm] {};
      \draw (t) edge node {$(\top,a)$} (p);
      \node[position] (q) [below left of = p,yshift=3mm] {};
      \draw (p) edge node[above left,xshift=2mm,yshift=-1mm] {$(0,b)$} (q);
      \node[position] (fu) [below of = q,yshift=3mm] {};
      \draw (q) edge node[left] {$\bot$} (fu);     
      \node[position] (r) [below right of = p,yshift=3mm] {};
      \draw (p) edge node[above right,xshift=-2mm,yshift=-1mm] {$(1,c)$} (r);
      \node[position] (s) [below left of = r,yshift=3mm] {};
      \draw (r) edge node[above left,xshift=2mm,yshift=-1mm] {$(0,d)$} (s);
      \node[position] (f') [below of = s,yshift=3mm] {};      
      \draw (s) edge node[left] {$\bot$} (f');
      \node[position] (t) [below right of = r,yshift=3mm] {};
      \draw (r) edge node[above right,xshift=-2mm,yshift=-1mm] {$(1,d)$} (t);
      \node[position] (f'') [below of = t,yshift=3mm] {};      
      \draw (t) edge node[left] {$\bot$} (f'');
    \end{scope}
  \end{tikzpicture}}
  \caption{A binary tree, and its encoding in $\fim(A')$.\label{fig.tree}}
  \label{fig.twa}
\end{figure}

As an example, consider the binary tree on the left of
Figure~\ref{fig.tree}. It is modelled by the following element of
$\fim(A')$, obtained from the Munn bi-rooted tree represented on the
right by a depth-first search:
$(\top,a) (0,b) \bot \ovl\bot \,\ovl{(0,b)} \,(1,c) (0,d) \bot
  \ovl\bot \,\ovl{(0,d)}\, (1,d) \bot \ovl \bot\, \ovl{(1,d)}\,
  \ovl{(1,c)} \,\ovl{(\top,a)}$.

When restricting the semantics of weighted automata and expressions to
elements of $\fim(A')$ that are encoding of trees,
Theorem~\ref{theo.eqatoe} gives an interesting model of weighted
expressions equivalent to weighted tree-walking automata over
rationally additive semirings.

The depth-first search of a tree is describable by an unambiguous weighted
expression (and thus also an unambiguous weighted automaton): letting
$(i,A)$ denote $\sum_{a\in A} (i,a)$, and restricting ourselves to
trees with nodes of arity 0 or 2 to simplify the writing, we let
\[W_0 = (0,A) ^* \cdot \bot\ , \quad
W_1 = \ovl{\bot} \cdot \ovl{(1,A)}^*\ , \quad\text{ and }\quad
W_{\text{succ}} = W_1\cdot \ovl{(0,A)} \cdot (1,A) \cdot W_0.\]
The weighted expression $W_0$ finds the leftmost leaf;
$W_1$ returns to the root from the rightmost leaf; and
$W_{\text{succ}}$ goes from a leaf to the next one
in the depth-first search. Then, the depth-first
search is implemented by the weighted expression
$(\top, A) \cdot W_0 \cdot W_{\text{succ}}^* \cdot W_1\cdot \ovl{(\top,A)}$.

By Theorem~\ref{theo.nonamb}, there exists an equivalent non ambiguous
automaton, that thus visits the whole tree. Since it is possible to
\emph{reset} the tree, going back to the root, in a non ambiguous
fashion, we can remove the requirement for the automata and the
expressions to visit the whole tree while starting and ending at the
root. This allows for more freedom in the models.

Taking advantage of this relaxation, it is possible to count the
maximal number of occurrences of a letter $a$ in branches of the tree,
starting at the root of the tree, non-deterministically going down the
chosen branch, and ending at the bottom: using the rationally additive
semiring $(\N\cup\{-\infty,+\infty\}, \sup, +, -\infty, 0)$,
\[\big((\top,a)\cdot 1 + (\top,A\setminus \{a\})\big)\cdot
  \big((0,a)\cdot 1+(0,A\setminus\{a\})+(1,a)\cdot
  1+(1,A\setminus\{a\})\big)^*\cdot \bot.\]

\section{Conclusion}
We have given an application of our result to tree-walking automata. A natural extension consists in investigating 
other kinds of structure like Mazurkiewicz traces or grids.

Our approach \emph{is} able to capture tree-walking automata, however
it is intrinsically more of a tree-\emph{generating} automaton
model. Over trees it does not make a huge difference but it does if we
try to extend this approach to more general graph-walking automata
models. A natural way to define weighted automata over graphs is to
take the sum of the weights of all paths over a given graph (in a
sense already explored in~\cite{M13}, but limiting itself to
non-looping runs). This means that a given path in the automaton can
be a run in different graphs, which is not compatible with our
approach of generating monoid elements.


One possible research direction would be to consider so-called SD-expressions introduced by Schützenberger (see \cite{DiekertW16}). These expressions were shown to coincide with star-free expressions with the advantage of not using the complement (instead restricting the languages over which the Kleene star can be applied, namely to prefix codes with bounded synchronisation delay) which means it can be applied to the quantitative setting. Indeed, in \cite{DartoisGK21}, the authors extended the result to transducers and showed that these expressions correspond to aperiodic transducers.
These expressions are naturally adapted to the unambiguous setting (maybe this restriction can be overcome)
but it would be interesting to study their expressive power in the context of pre-rational monoids.

A final direction would be to use logics instead of expressions, to
describe in a less operational way the behaviour of weighted automata
over monoids. Promising results have already been obtained in specific
contexts, like non-looping automata walking (with pebbles) on words,
trees or graphs~\cite{BGMZ14a}, but a cohesive point of view via
monoids is still lacking.

\bibliographystyle{plainurl}
\bibliography{generalised}


\newpage

\appendix

\section{Non-trivial example of pre-rational monoid}\label{app:prerat}

We explain why the monoid
$(M=\{L \subseteq A^* \mid \varepsilon\in L\}, \cdot,
\{\varepsilon\})$, with $A$ a finite alphabet, is pre-rational. Notice
that it is not finitely generated, nor fulfils the condition of
Lemma~\ref{lem.pref}. We thus follow the definition of
pre-rationality. Consider thus a finite alphabet $B$, a morphism
$\mu\colon B^* \to M$ and a language $L\in M$. To show that
$\mu^{-1}(L)$ is a rational language of $B^*$, consider the languages
\[U_L=\{w\in B^* \mid L\subseteq \mu(w)\} \qquad \text{ and } \qquad
  O_L=\{w\in B^* \mid \mu(w)\not \subseteq L\}\] Then, $\mu^{-1}(L) = U_L \cap
(B^* \setminus O_L)$, and it is sufficient to show that both $U_L$ and
$O_L$ are rational languages to conclude.

Since all languages of $M$ contain the empty word, the languages $U_L$
and $O_L$ are upward-closed with respect to the subsequence relation:
for all $u,v,w\in B^*$, if $uv\in U_L$ (respectively, $uv\in O_L$)
then $uwv\in U_L$ (respectively, $uwv\in O_L$). By Higman's lemma,
showing that the subsequence relation is a well quasi-order, $U_L$ and
$O_L$ have a finite set of minimal elements. Then, the two sets can be
recovered from the minimal elements $b_0b_1\cdots b_{n-1}$ by adding
$A^*$ in-between each letter, i.e.~considering the piecewise-testable
languages $B^* b_0 B^* b_1 B^*\cdots B^* b_{n-1}B^*$: $U_L$ and $O_L$
are thus finite unions of rational languages, therefore rational.

\end{document}